\documentclass{aa}
\usepackage{psfig}
\usepackage{times}
\def\mag{\hbox{$^{\rm m}$}}
\begin{document}
\thesaurus{08                 
           (08.03.4;          
            08.05.2;          
            08.09.2: AB Aur, HD 163296;  
            08.16.5;          
            13.09.6)}         
%

\title{ISO spectroscopy of circumstellar dust in the Herbig Ae systems AB Aur and HD 163296
\thanks{Based on observations with ISO, an ESA project with instruments
        funded by ESA Member States (especially the PI countries: France,
        Germany, the Netherlands and the United Kingdom) and with the
        participation of ISAS and NASA and on observations collected at 
        the European Southern Observatory, La Silla, Chile.}}

\author{M.E. van den Ancker\inst{1,2} \and J. Bouwman\inst{1} \and 
        P.R. Wesselius\inst{3} \and L.B.F.M. Waters\inst{1,4} \and 
        S.M. Dougherty\inst{5,6} \and E.F. van Dishoeck\inst{7}}

\institute{Astronomical Institute ``Anton Pannekoek'', University of
 Amsterdam, Kruislaan 403, NL--1098 SJ Amsterdam, The Netherlands
\and Harvard-Smithsonian Center for Astrophysics, 60 Garden Street, MS 42, 
 Cambridge  MA 02138, USA 
\and SRON Laboratory for Space Research Groningen, P.O. Box 800,
 NL--9700 AV Groningen, The Netherlands 
\and Instituut voor Sterrenkunde, Katholieke Universiteit Leuven, 
 Celestijnenlaan 200B, B-3001 Heverlee, Belgium
\and Dept. of Physics \& Astronomy, University of Calgary, 2500 University 
 Drive NW, Calgary, Alberta T2N 1N4, Canada 
\and Dominion Radio Astrophysical Observatory, P.O. Box 248, White Lake Road, 
 Penticton, British Columbia V2A 6K3, Canada
\and Leiden Observatory, P.O. Box 9513, NL--2300 RA Leiden, The Netherlands}

\offprints{M.E. van den Ancker (mario@astro.uva.nl)}
\date{Received <date>; accepted <date>}

\maketitle

\begin{abstract}
Using both the Short- and Long-wavelength Spectrometers on board the 
Infrared Space Observatory (ISO), we have obtained infrared spectra of 
the Herbig Ae systems AB~Aur and HD~163296. In addition, we 
obtained ground-based $N$ band images of HD~163296. Our results 
can be summarized as follows: 
(1) The main dust components in AB~Aur are amorphous silicates, iron oxide 
    and PAHs; 
(2) The circumstellar dust in HD~163296 consists of amorphous silicates, 
    iron oxide, water ice and a small fraction of crystalline silicates; 
(3) The infrared fluxes of HD~163296 are dominated by solid state features; 
(4) The colour temperature of the underlying continuum is much cooler 
    in HD~163296 than in AB~Aur, pointing to the existence of a 
    population of very large (mm sized) dust grains in HD~163296; 
(5) The composition and degree of crystallization of circumstellar 
    dust are poorly correlated with the age of the central star. The 
    processes of crystallization and grain growth are also not 
    necessarily coupled. This means that either the evolution of 
    circumstellar dust in protoplanetary disks happens very rapidly 
    (within a few Myr), or that this evolution is governed by factors 
    other than stellar mass and age.
\keywords{Circumstellar matter -- Emission-line stars -- Stars: AB Aur, 
          HD 163296 -- Pre-main sequence stars -- Infrared: Stars}
\end{abstract}

\section{Introduction}
Herbig Ae/Be stars are intermediate-mass pre-main sequence stars 
surrounded by disks of gas and dust which might be the 
site of on-going planet formation (see Waters \& Waelkens 1998 
for a recent review). AB~Aur (A0Ve) and HD~163296 (A3Ve) belong to 
the best studied Herbig stars and are sometimes considered prototypical 
for the entire class. As early as 1933, Merrill \& Burwell 
remarked upon the similarity of both systems, which was confirmed 
by numerous subsequent authors. Apart from the fact that the 
stellar mass, effective temperature and age of AB~Aur and 
HD~163296 are nearly identical (van den Ancker et al. 1997, 1998), 
the similarity between the two systems also extends to their 
circumstellar environment: both AB~Aur and HD~163296 are examples 
of relatively isolated star formation, and are not hindered by 
confusion with other sources (Henning et al. 1998; Di Francesco et al. 1998).

Both AB~Aur and HD~163296 show a rich, variable, emission-line spectrum 
from the ultraviolet to the optical. With a few exceptions, most 
noticeably the observed [O\,{\sc i}] emission, these lines have 
been successfully modelled as arising in an inhomogeneous stellar 
wind (Catala et al. 1989; B\"ohm et al. 1996; Bouret et al. 1997).
Infall of material has been detected in both HD~163296 and AB~Aur 
through monitoring of UV and optical absorption and emission lines 
and can be explained by the presence of infalling evaporating 
exocomets (Grady et al. 1996, 1999). In the infrared, both stars 
are among the sources with the strongest 10~$\mu$m silicate feature 
in emission (Cohen 1980; Sitko 1981; Sorrell 1990). Sitko et al. 
(1999) compared the 10~$\mu$m silicate feature in HD~163296 with 
that of solar-system comets and found a striking resemblance 
with that of comet Hale-Bopp. Neither star shows the 3.29~$\mu$m UIR 
emission band present toward many Herbig Ae/Be stars 
(Brooke et al. 1993; Sitko et al. 1999).
Basic astrophysical parameters of both stars are listed in Table~1.

There is strong evidence for the presence of a circumstellar disk 
in both AB~Aur and HD~163296. Bjorkman et al. (1995) detected a 
90\degr~flip of the polarization angle between the optical and the 
ultraviolet in HD~163296, which they interpreted as evidence for 
a flattened, disk-like structure. Mannings \& Sargent (1997) 
resolved the gaseous disks surrounding AB~Aur and HD~163296 using 
CO millimeter wave aperture synthesis imaging. 
Using continuum measurements at 1.3~mm, the 
same authors also resolved the circumstellar dust disk of HD~163296. 
The AB~Aur dust disk has also been resolved in the infrared, and 
shows a surprisingly strong dependence of disk diameter 
on wavelength, ranging from 0\farcs0065 (0.94~AU) at 2.2~$\mu$m 
(Millan-Gabet et al. 1999), through 0\farcs24 (35 AU) at 11.7~$\mu$m 
to 0\farcs49 (70~AU) at 17.9~$\mu$m (Marsh et al. 1995).

In this paper we present new infrared spectra of AB~Aur and 
HD~163296 obtained with the Short- and Long Wavelength Spectrometers 
on board the {\it Infrared Space Observatory} (ISO; Kessler et al. 1996). 
We will discuss these spectra and their implications for the evolution 
of dust in Herbig systems. In a subsequent paper (Bouwman et al. 2000), 
we will describe a model for the circumstellar dust disks of AB~Aur and 
HD~163296 and apply it to these data.

\section{Observations}
ISO Short Wavelength (2.4--45~$\mu$m) Spectrometer (SWS; de Graauw et al. 
1996) and Long Wavelength (43--197~$\mu$m) Spectrometer (LWS; Clegg et al. 
1996) full grating scans of AB~Aur were obtained in ISO revolutions 680 
(at JD 2450717.747) and 835 (JD 2450872.380), respectively. An SWS full 
grating scan of HD~163296 was made in revolution 329 (JD~2450367.398). 
Observing times were 3666 seconds for the SWS and 2741 seconds for the 
LWS observations. Data were reduced in a standard fashion using calibration 
files corresponding to OLP version 7.0 (SWS) or 6.0 (LWS), after 
which they were corrected for remaining fringing and glitches. To 
increase the S/N in the final spectra, statistical outliers were removed 
and the detectors were aligned, after which the spectra were rebinned 
to a lower spectral resolution. The resulting spectra are shown in Fig.~1.

$N$-band (10.1~$\mu$m) images of HD~163296 were obtained on July 25, 
1997 (at JD 2450654.608) using TIMMI on the ESO 3.6m telescope at La Silla. 
Total integration time was 65 minutes. The pixel size was 0\farcs336, 
with a total field of view of 21\farcs5~$\times$~21\farcs5. 
After a standard reduction procedure, the HD~163296 image was 
indistinguishable from that of the 
standard star $\eta$~Sgr. After deconvolution of the HD~163296 image 
with that of $\eta$~Sgr, the resulting stellar image stretched across 
2 pixels. We conclude that the bulk of the 10~$\mu$m flux of HD~163296 
comes from an area less than 0\farcs7 (90~AU at 122~pc) in diameter.

\section{Contents of spectra}
The infrared spectra of AB~Aur and HD~163296 (Fig.~1) show big 
differences: whereas AB~Aur shows the cool, strong 
continuum expected for a Herbig star, for HD~163296 the continuum 
appears to be so weak that the entire 
SWS spectrum is dominated by solid-state emission features. The 
IRAS fluxes also plotted in Fig.~1 suggest that an underlying 
continuum in HD~163296 is present, but peaks longward of 
100~$\mu$m and therefore is much cooler than the $\approx$40~K 
continuum in AB~Aur.
\begin{table}
\caption[]{Astrophysical parameters of programme stars}
\begin{flushleft}
\begin{tabular}{lcccc}
\hline\noalign{\smallskip}
                       & AB Aur           & Ref. & HD 163296        & Ref.\\
\noalign{\smallskip}
\hline\noalign{\smallskip}
$\alpha$ (2000)        & 04 55 45.79       & (1) & 17 56 21.26       & (1)\\
$\delta$ (2000)        & +30 33 05.5       & (1) & $-$21 57 19.5     & (1)\\
$d$ [pc]               & 144$^{+23}_{-17}$ & (2) & 122$^{+17}_{-13}$ & (2)\\
Sp. Type               & A0Ve+sh           & (3) & A3Ve              & (8)\\
$V$ [\mag]             & 7.03--7.09$^\dagger$&(2)& 6.82--6.89        & (2)\\
$E(B-V)$ [\mag]        & 0.16$\pm$0.02     & (4) & 0.04$\pm$0.02     & (4)\\
$T_{\rm eff}$ [K]      & 9500$^{+500}_{-300}$&(2)& 8700$^{+200}_{-300}$& (4)\\
$L_\star$ [L$_\odot$]  & 47$\pm$12         & (2) & 26$\pm$5          & (4)\\
$R_\star$ [R$_\odot$]  & 2.5$\pm$0.5       & (4) & 2.2$\pm$0.2       & (4)\\
$M_\star$ [M$_\odot$]  & 2.4$\pm$0.2       & (2) & 2.0$\pm$0.2       & (4)\\
log(Age) [yr]          & 6.3$\pm$0.2       & (2) & 6.6$\pm$0.2       & (4)\\
H$\alpha$              & P Cygni           & (5) & Double-peaked     & (9)\\
$v \sin i$ [km s$^{-1}$]& 80$\pm$5         & (5) & 120$\pm$1         & (10)\\
rot. period [$^{\rm h}$]& 32$\pm$3         & (6) & 35$\pm$5          & (11)\\
$i$ [\degr]            & 76                & (7) & 58                & (7)\\
P.A. [\degr]           & +79$^{+2}_{-3}$   & (7) & +126$^{+2}_{-3}$  & (7)\\
\noalign{\smallskip}
\hline
\end{tabular}
\end{flushleft}
$^\dagger$However, older photographic measurements (Gaposchkin 1952) show values up to $m_{pg}=8.4$ for AB Aur.\medskip\\
\noindent
References: 
(1) Hipparcos Catalogue; 
(2) van den Ancker et al. (1998); 
(3) B\"ohm \& Catala (1993); 
(4) This paper; 
(5) B\"ohm \& Catala (1995); 
(6) B\"ohm et al. (1996); 
(7) Mannings \& Sargent (1997); 
(8) Gray \& Corbally (1998); 
(9) Pogodin (1994); 
(10) Halbedel (1996); 
(11) Catala et al. (1989).
\end{table}

Both AB~Aur and HD~163296 show a strong 9.7~$\mu$m amorphous 
silicate feature in emission together with a broad emission 
complex ranging from 14 to 38~$\mu$m. The emissivities for various 
dust components in the spectra are also included in Fig.~1. 
In the HD~163296 spectrum the broad emission complex 
ranging from 14 to 38~$\mu$m is too broad and too intense to be 
solely attributed to the 19~$\mu$m feature due to the O--Si--O 
bending mode. We tentatively attribute this feature to a blend 
of silicates and iron oxide. Lab spectra of FeO also show a 
strongly rising emissivity in the short-wavelength 
range of the SWS (Henning et al. 1995). When folded with a 
$\approx$~800~K blackbody, this naturally produces a broad 
emission feature peaking around 3~microns, which is present in 
both stars.

The large degree of redundancy in the SWS data makes it possible to 
assess the reality of weak spectral features which at first glance may 
appear to be lost in the noise. Each part of the spectrum was scanned 
twice by twelve detectors, so by checking whether a particular feature 
is seen in all detectors and in both scan directions, it is possible to 
disentangle real features from noise. The 
features identified in this way are listed in Table~2. AB~Aur clearly 
shows the familiar 6.2 and 7.7 and possibly also the 8.6 and 11.2~$\mu$m 
UIR bands usually attributed to emission by polycyclic aromatic 
hydrocarbons (PAHs), as well as a new UIR band at 15.9~$\mu$m. The 
3.29~$\mu$m UIR band is absent.
\begin{figure}[t]
\vspace{0.1cm}
\centerline{\psfig{figure=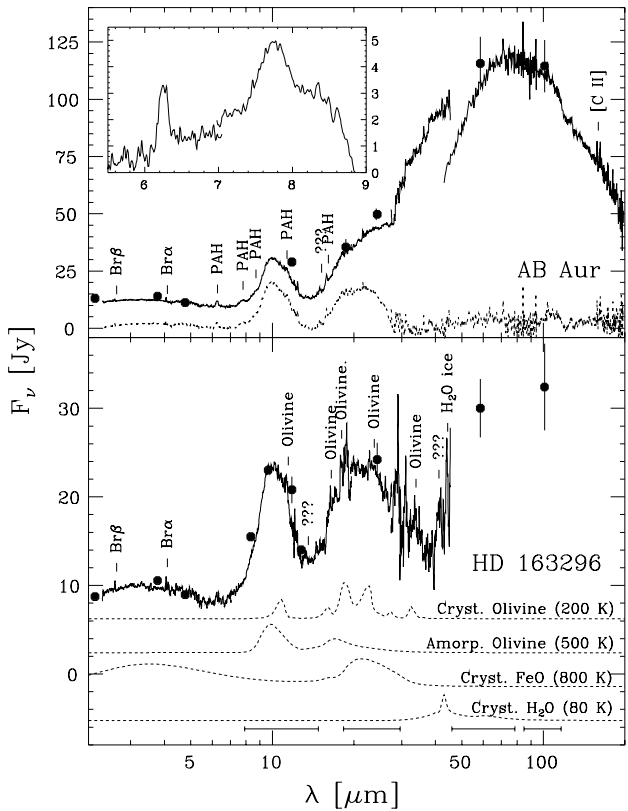,width=8.2cm,angle=0}}
\caption[]{ISO spectra of AB Aur (top) and HD 163296 (bottom) 
           with the main features identified. The solid dots indicate 
           ground-based and IRAS photometry (Berrilli et al. 1987, 1992; 
           Hillenbrand et al. 1992; Weaver \& Jones 1992). The dashed line 
           in the top plot is the AB Aur spectrum after subtracting a spline 
           fit to the continuum. 
           Also shown (dashed lines at bottom) are the emissivities 
           at a given temperature of the various dust components in 
           the spectra (J\"ager et al. 1998; Dorschner et al. 1995; 
           Henning et al. 1995; Bertie et al. 1969). The bars at the 
           bottom of the plot show the IRAS 12, 25, 60 and 100~$\mu$m 
           passbands (FWHM). The inset in the AB Aur plot shows the UIR 
           bands in the 5.5--9.0~$\mu$m region after subtracting a 
           pseudo-continuum.}
\end{figure}
\begin{figure}[t]
\vspace{0.1cm}
\centerline{\psfig{figure=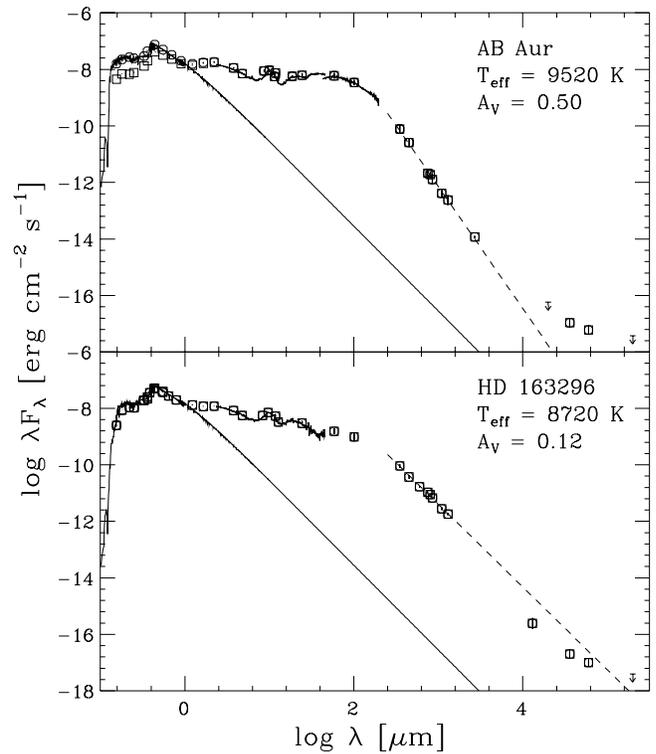,width=8.45cm,angle=0}}
\caption[]{Spectral energy distributions of AB Aur (top) and HD 163296 (bottom). 
           Open squares and circles indicate observed and extinction corrected fluxes 
           from literature (Wesselius et al. 1982; Th\'e et al. 1985; 
           Strom et al. 1972; Cohen \& Schwartz 1976; Berrilli et al. 1987, 1992; 
           Hillenbrand et al. 1992; Weaver \& Jones 1992; Mannings 1994; 
           G\"udel et al. 1989; Brown et al. 1993; Skinner et al. 1993). The 
           solid lines show Kurucz models for the stellar photospheres of AB Aur 
           and HD 163296. The dashed lines show linear fits to the submm 
           data points for comparison with the slope of the Rayleigh-Jeans 
           tail of the Kurucz model.}
\end{figure}

The HD~163296 spectrum shows a number of small emission features 
at wavelengths corresponding to those of crystalline olivines 
((Mg$_x$Fe$_{1-x}$)$_2$SiO$_4$). Remarkably, the amorphous 
silicate feature has a higher band-strength relative to the 
continuum in our ISO data than in the ground-based 8--13~$\mu$m 
spectrum of HD~163296 by Sitko et al. (1999). In contrast, the 
11.2~$\mu$m shoulder, due to crystalline silicates, appears 
weaker in our spectrum, suggesting a significant time variability 
of both components.

In addition to this, HD~163296 shows emission from the 44~$\mu$m H$_2$O 
ice feature. The relative location of the IRAS 60~$\mu$m measurement 
in comparison to the SWS spectrum suggests that the long-wavelength 
H$_2$O ice feature around 69~$\mu$m as well as the broad unidentified 
feature longward of 100~$\mu$m, observed in HD~100546 and HD~142527 
(Malfait et al. 1998, 1999), might also be very prominent in HD~163296.
PAHs are very weak or absent in HD~163296.

In addition to the solid-state features, both AB~Aur and HD~163296 also 
contain a number of  H\,{\sc i} recombination lines at shorter wavelengths. 
All lines from the Bracket and Pfund included in the SWS wavelength 
range series are present, while the higher H\,{\sc i} series are not 
detected, possibly due to the combined effect of a lower instrumental 
sensitivity and a higher background in this part of the spectrum. These 
recombination line data will be discussed in more detail in a forthcoming 
paper.

The LWS spectrum of AB~Aur is relatively smooth and featureless. 
The only line that is clearly visible in the spectrum is the 
[C\,{\sc ii}] line at 157.7~$\mu$m. The strength of this line 
(8.6 $\times$ 10$^{-16}$~W~m$^{-2}$) is compatible with it 
originating in the background rather than being circumstellar. 
As can be seen from Fig.~1, there is a $\approx$25\% difference 
in the flux scales between the AB~Aur SWS and LWS spectra in 
the overlapping region. Although within the formal errors of 
the absolute flux calibration for SWS and LWS, this discrepancy 
is larger than that found in other sources. The difference 
cannot be attributed to the different aperture sizes 
(33\arcsec~$\times$ 20\arcsec~for SWS versus a circular 
80\arcsec~FWHM for LWS), and confusion with extended emission, 
since then the LWS spectrum would have to have a higher flux 
level than the SWS spectrum. It is interesting to note that 
in the time interval between the SWS and LWS measurements, 
AB Aur did show an optical photometric event which could have also 
affected the infrared brightness (van den Ancker et al. 1999 and 
references therein), so the difference in flux between SWS and 
LWS might in fact be due to real variability. In Fig.~1 we also 
plotted the IRAS fluxes of AB~Aur. Although compatible with both 
spectra, the IRAS 60~$\mu$m flux agrees better with the SWS 
spectrum, so we rather arbitrarily choose to adopt the SWS flux 
calibration for the region around 45~$\mu$m.
\begin{table*}
\caption[]{Contents of Spectra}
\begin{flushleft}
\begin{tabular}{lccccccccccccccc}
\hline\noalign{\smallskip}
Feature           & FeO   & PAH    & PAH    & PAH   & PAH   & Si-O  & PAH    &        & PAH    & O-Si-O & FeO\\
$\lambda$ [$\mu$m]& [3]   & [3.3]  & [6.2]  & [7.7] & [8.6] & [9.7] & [11.2] & [15.0] & [15.9] & [19]   & [23]\\
\noalign{\smallskip}
\hline\noalign{\smallskip}
AB Aur            &$\surd$& --     &$\surd$ &$\surd$&$\surd$&$\surd$&$\surd$:&$\surd$:&$\surd$&$\surd$  &$\surd$\\
HD 163296         &$\surd$&$\surd$:&$\surd$:& --    & --    &$\surd$& --     & --     & --    &$\surd$  &$\surd$\\
\noalign{\smallskip}
\hline
\noalign{\vspace*{0.45cm}}
\hline\noalign{\smallskip}
Feature           & Oliv.  &        & Oliv.  & Oliv.  & Pyr.?  & Oliv.  & Oliv.  &  Oliv. &        & H$_2$O ice \\
$\lambda$ [$\mu$m]& [11.3] & [13.4] & [16.3] & [17.8] & [18.2] & [23.5] & [31.3] & [33.5] & [40.7] & [43.8]\\
\noalign{\smallskip}
\hline\noalign{\smallskip}
AB Aur            & --     & --     & --     & --     & --     & --     &$\surd$:&$\surd$:&$\surd$:& --\\
HD 163296         &$\surd$ &$\surd$ &$\surd$ &$\surd$ &$\surd$ &$\surd$ &$\surd$ &$\surd$ &$\surd$ &$\surd$\\
\noalign{\smallskip}
\hline
\end{tabular}
\end{flushleft}
\end{table*}

\section{Spectral energy distributions}
Spectral Energy Distributions (SEDs) of AB~Aur and HD~163296 were 
constructed from literature data as well as our new ISO spectra and 
newly obtained VLA photometry for HD~163296 and are shown in Fig.~2. 
All the submm fluxes in 
these SEDs refer to single-dish measurements. As can be seen from 
Fig.~2, the SED can be naturally decomposed in three parts: the 
optical wavelength range, where the total system flux is dominated 
by the stellar photosphere, the infrared to submm, where emission 
originates from the circumstellar dust disk, and the radio, where 
free-free emission from the stellar wind becomes dominant.

The difference in behaviour of the dust component in HD~163296 and 
AB~Aur is striking: after a nearly flat energy distribution 
in the infrared, the sub-mm and mm fluxes of AB~Aur drop rapidly 
($\lambda F_\lambda \propto \lambda^{-4.3}$), indicative of the dust 
becoming optically thin at these wavelengths, whereas toward HD~163296 
the slope of the sub-mm to mm fluxes ($\propto \lambda^{-2.9}$)
is within errors equal to that of the Rayleigh-Jeans tail of a 
black body ($\propto \lambda^{-3}$). The new radio points at 1.3, 
3.6 and 6~cm for HD~163296 do not follow the simple power-law 
dependence expected if these were solely due to free-free radiation. 
This demonstrates that even at wavelengths as long as 1.3~cm, a 
significant fraction of the system flux is due to circumstellar 
dust. The 3.6 and 6~cm fluxes are probably dominated by free-free 
emission.

The energy distribution of a circumstellar dust disk 
is governed by its temperature profile, the density distribution 
and the dust properties (chemical composition and size distribution). 
Since the circumstellar disks of AB~Aur and HD~163296 stars are 
expected to be passive (Waters \& Waelkens 1998) and the properties 
of the central stars are nearly identical, the temperature profiles 
in the disks are expected to be similar as well. One possibility to 
explain the different sub-mm to mm slope for AB~Aur and HD~163296 
could be a much flatter density distribution for HD~163296. However, 
with a standard sub-mm dust emissivity ($\beta$ = 2) the inferred 
dust mass for HD~163296 would become implausibly large. A better 
explanation may be that the dust properties of AB~Aur and 
HD~163296 are different, a fact already concluded independently 
from the ISO spectra. To be able to radiate efficiently, dust 
particles must have a size similar to (or larger than) the 
wavelength, pointing to the existence of a population of 
mm- to cm-sized cold dust grains in the circumstellar environment 
of HD~163296, whereas those in AB~Aur must be micron-sized. The ISO 
spectrum of HD~163296 also contains warm ($\approx$ 800~K) dust, 
suggesting a significant lack of emission from dust of intermediate 
temperatures.

\section{Discussion and conclusions}
We have shown that the main difference between the AB~Aur and 
HD~163296 systems is that HD~163296 contains a population of very 
large (mm to cm-sized), cold, partially crystalline, dust grains, 
which is absent in AB~Aur. AB~Aur contains a population of small 
dust grains (PAHs), which is absent in HD~163296. In view of the fact 
that the stellar parameters are nearly identical except for stellar 
rotation, these differences are remarkable. This must mean that either 
the evolution of the dust composition in protoplanetary disks happens 
within the error in the age determination of both systems 
(2$^{+2}_{-1}$ Myr for AB~Aur vs. 4$^{+4}_{-2}$ Myr for HD~163296; 
van den Ancker et al. 1998), or that the evolution of the dust 
is dominated by external factors.

We have also shown that in AB~Aur and HD~163296 iron oxide is a  
constituent of the circumstellar dust mixture and can also be 
responsible for the observed excess emission near 3~$\mu$m. 
Since the near-infrared excess exhibited by our programme stars 
is by no means unusual for a Herbig Ae/Be star, this means that 
the same applies for the entire group of Herbig stars. Therefore 
the results of models attributing this near infrared excess emission 
to very hot dust from an actively accreting disk (e.g. Hillenbrand 
et al. 1992) must be regarded with some caution. The 
case of HD~163296 demonstrates that infrared broad-band photometry 
can be completely dominated by emission from solid-state features, 
which must also be taken into account in any future modelling of 
the energy distributions of Herbig stars.

The detection of PAHs in AB~Aur shows that ground-based 
surveys (e.g. Brooke et al. 1993) have underestimated the 
fraction of Herbig stars containing PAHs. The case of HD~163296 
shows that the fraction of Herbig stars showing PAH emission will 
not go up to 100\%, so models depending on the presence of very 
small dust grains to explain the observed near-infrared excess 
in Herbig Ae/Be stars (Natta et al. 1993; Natta \& Kr\"ugel 1995) 
will not be successful in all cases. We believe iron oxide 
to be a more plausible explanation for this near-infrared excess.

It is difficult to attribute the absence of the 3.29~$\mu$m 
feature in AB~Aur to an unusual temperature of the PAHs. Since 
the 6.2 and 7.7~$\mu$m C--C stretches are strong, while the 
bands due to C--H bonds are weak or absent, a more 
promising possibility seems a very low hydrogen covering factor 
of the PAHs in AB~Aur or the presence of a population of large 
($>$ 100 C atoms) PAHs (Schutte et al. 1993). If the new 
15.9~$\mu$m UIR band in AB~Aur is also caused by PAHs, this 
suggests that it is also due to a C--C bond. 

To gain more insight in the evolution of dust in protoplanetary 
disks, it is useful to compare the spectra presented here to those 
of other Herbig Ae stars. Except for the absence of crystalline 
material, the AB~Aur spectrum and energy distribution are nearly 
identical to those of its older counterpart HD~100546 (Malfait et al. 
1998). In the case of AB~Aur (2.5~M$_\odot$), any possible 
crystallization of circumstellar dust must therefore occur at a 
stellar age older than 2 $\times$ 10$^6$ years. 
The differences between HD~100546 and HD~163296 are larger: 
the crystalline dust in HD~100546 is much more prominent than that 
in HD~163296 and the population of large cold dust grains seen in 
HD~163296 is absent in HD~100546. 

As these cases of AB~Aur, HD~163296 and HD~100546 demonstrate, the 
age of the central star and the degree of crystallization do not 
show a one to one correspondence, 
but the processes of grain growth and crystallization in protoplanetary 
disks are also not necessarily coupled. Studying a larger sample of 
Herbig Ae/Be stars might shed more light on what causes this large 
observed diversity in dust properties in systems which appear very 
similar in other aspects.

\acknowledgements{This paper is 
based on observations with ISO, an ESA project with instruments
funded by ESA Member States (especially the PI countries: France,
Germany, the Netherlands and the United Kingdom) and with the
participation of ISAS and NASA and on observations collected at 
the European Southern Observatory, La Silla, Chile. 
The authors would like to thank the SWS IDT for their help with the 
SWS observations and Norman Trams and Michelle 
Creech-Eakman for their help with the LWS observations. Koen 
Malfait and Xander Tielens are gratefully acknowledged for reading 
of the manuscript of this paper prior to publication.}


\begin{thebibliography}{}
\bibitem{}
Berrilli, F., Corciulo, G., Ingrosso, G., Lorenzetti, D., Nisini, B., 
 Strafella, F. 1993, ApJ 398, 254
\bibitem{}
Berrilli, F., Lorenzetti, D., Saraceno, P., Strafella, F. 1987, MNRAS 228, 833
\bibitem{}
Bertie, J.B. Labb\'e, H.J., Whalley, B. 1969, J. Chem. Phys. 50, 4501
\bibitem{}
Bjorkman, K.S. et al. 1995, BAAS 27, 1319
\bibitem{}
B\"ohm, T., Catala, C. 1993, A\&AS 101, 629
\bibitem{}
B\"ohm, T., Catala, C. 1995, A\&A 301, 155
\bibitem{}
B\"ohm, T., Catala, C., Donati, J.F. et al. 1996, A\&AS 120, 431
\bibitem{}
Bouret, J.C., Catala, C., Simon, T. 1997, A\&A 328, 606
\bibitem{}
Brooke, T.Y., Tokunaga, A.T., Strom, S.E. 1993, AJ 106, 656
\bibitem{}
Brown, D.A., P\'erez, M.R., Yusef-Zadeh, F. 1993, AJ 106, 2000
\bibitem{}
Catala, C., Simon, T., Praderie, F., Talavera, A., Th\'e, P.S., 
 Tjin A Djie, H.R.E. 1989, A\&A 221, 273
\bibitem{}
Clegg, P.E., Ade, P.A.R. et al. 1996, A\&A 315, L38
\bibitem{}
Cohen, M. 1980, MNRAS 191, 499
\bibitem{}
Cohen, M., Schwartz, R.D. 1976, MNRAS 174, 137
\bibitem{}
de Graauw, Th., Haser, L.N. et al. 1996, A\&A 315, L49
\bibitem{}
Di Francesco, J., Evans, N.J., Harvey, P.M., Mundy, L.G., Butner, H.M. 
 1998, ApJ 509, 324
\bibitem{}
Dorschner, J., Begemann, B., Henning, Th., J\"ager, C., Mutschke, H. 
 1995, A\&A 300, 503
\bibitem{}
Grady, C.A., P\'erez, M.R., Talavera, A. et al. 1996, A\&AS 120, 157
\bibitem{}
Grady, C.A., P\'erez, M.R., Bjorkman, K.S., Massa, D. 1999, ApJ 511, 925
\bibitem{}
Gray, R.O., Corbally, C.J. 1998, AJ 116, 2530
\bibitem{}
G\"udel, M., Benz, A.O., Catala, C., Praderie, F. 1989, A\&A 217, L9
\bibitem{}
Halbedel, E.M. 1996, PASP 108, 833
\bibitem{}
Henning, Th., Begemann, B., Mutschke, H., Dorschner, J. 1995, A\&AS 112, 143
\bibitem{}
Henning, Th., Burkert, A., Launhardt, R., Leinert, C., Stecklum, B. 1998, 
 A\&A 336, 565
\bibitem{}
Hillenbrand, L.A., Strom, S.E., Vrba, F.J., Keene, J. 1992, ApJ 397, 613
\bibitem{}
J\"ager, C., Molster, F.J., Dorschner, J., Henning, Th., Mutschke, H., 
 Waters, L.B.F.M. 1998, A\&A 339, 904
\bibitem{}
Kessler, M.F., Steinz, J.A., Anderegg, M.E., Clavel, J., et al.
 1996, A\&A 315, L27
\bibitem{}
Malfait, K., Waelkens, C., Waters, L.B.F.M., Vandenbussche, B., Huygen, E., 
 de Graauw, M.S. 1998, A\&A 332, L25
\bibitem{}
Malfait, K., Waelkens, C., Bouwman, J., de Koter, A., Waters, L.B.F.M. 
 1999, A\&A 345, 181
\bibitem{}
Mannings, V. 1994, MNRAS 271, 587
\bibitem{}
Mannings, V., Sargent, A.I. 1997, ApJ 490, 792
\bibitem{}
Marsh, K.A, van Cleve, J.E., Mahoney, M.J., Hayward, T.L., Houck, J.R. 
 1995, ApJ 451, 777
\bibitem{}
Merrill, P.W., Burwell, C.G. 1933, ApJ 78, 87
\bibitem{}
Millan-Gabet, R., Schloerb, F.P., Traub, W.A., Malbet, F., Berger, J.P., 
 Bregman, J.D. 1999, ApJ 513, L131
\bibitem{}
Natta, A., Kr\"ugel, E. 1995, A\&A 302, 849
\bibitem{}
Natta, A., Prusti, T., Kr\"ugel, E. 1993, A\&A 275, 527
\bibitem{}
Pogodin, M.A. 1994, A\&A 282, 141
\bibitem{}
Schutte, W.A., Tielens, A.G.G.M., Allamandola, L.J. 1993, ApJ 415, 397
\bibitem{}
Sitko, M.L. 1981, ApJ 247, 1024
\bibitem{}
Sitko, M.L., Grady, C., Lynch, D.K., Russell, R.W., Hanner, M.S. 
 1999, ApJ 510, 408
\bibitem{}
Skinner, S.L., Brown, A., Stewart, A.T. 1993, ApJS 87, 217
\bibitem{}
Sorrell, W.H. 1990, ApJ 361, 150
\bibitem{}
Strom, S.E., Strom, K.E., Yost, J., Carrasco, L., Grasdalen, G. 1972, ApJ 173, 353
\bibitem{}
Th\'e, P.S., Felenbok, P., Cuypers, H., Tjin A Djie, H.R.E. 1985, A\&A 149, 429
\bibitem{}
van den Ancker, M.E., Th\'e, P.S., Tjin A Djie, H.R.E., Catala, C., 
 de Winter, D., Blondel, P.F.C., Waters, L.B.F.M. 1997, A\&A 324, L33
\bibitem{}
van den Ancker, M.E., de Winter, D., Tjin A Djie, H.R.E. 1998, A\&A 330, 145
\bibitem{}
van den Ancker, M.E., Volp, A.W., P\'erez, M.R., de Winter, D. 1999, 
 Inf. Bull. Var. Stars 4704, 1
\bibitem{}
Waters, L.B.F.M., Waelkens, C. 1998, ARA\&A 36, 233
\bibitem{}
Weaver, W.B., Jones, G. 1992, ApJS 78, 239
\bibitem{}
Wesselius, P.R., van Duinen, R.J., de Jonge, A.R.W., Aalders, J.W.G., Luinge,
 W., Wildeman, K.J. 1982, A\&AS 49, 427
\end{thebibliography}
\end{document}